\documentclass{elsart}

\usepackage{amssymb}
\usepackage{mathpazo}

\usepackage{graphicx}
\newcommand{\ignore}[1]{}
\newcommand{\be}{\begin{equation}} \newcommand{\ee}{\end{equation}}
\newcommand{\ba}{\begin{eqnarray}} \newcommand{\ea}{\end{eqnarray}}
\newcommand{\nn}{\nonumber} \renewcommand{\bf}{\textbf}
\newcommand{\ra}{\rightarrow}

\renewcommand{\a}{\alpha} \renewcommand{\b}{\beta}

\def\slasha#1{\setbox0=\hbox{$#1$}#1\hskip-\wd0\hbox to\wd0{\hss\sl/\/\hss}}
\def\slashb#1{\setbox0=\hbox{$#1$}#1\hskip-\wd0\dimen0=5pt\advance
       \dimen0 by-\ht0\advance\dimen0 by\dp0\lower0.5\dimen0\hbox
         to\wd0{\hss\sl/\/\hss}}
\def\bra#1{\left< #1\right|}
\def\ket#1{\left| #1\right>}
\def\bracket#1#2{\left<#1\mid #2\right>}
\def\EV#1#2#3{\bra{#1}#2\ket{#3}}
\input{epsf}

\begin{document}

\begin{frontmatter}

% Title, authors and addresses

% use the thanksref command within \title, \author or \address for footnotes;
% use the corauthref command within \author for corresponding author footnotes;
% use the ead command for the email address,
% and the form \ead[url] for the home page:
% \title{Title\thanksref{label1}}
% \thanks[label1]{}
% \author{Name\corauthref{cor1}\thanksref{label2}}
% \ead{email address}
% \ead[url]{home page}
% \thanks[label2]{}
% \corauth[cor1]{}
% \address{Address\thanksref{label3}}
% \thanks[label3]{}

\title{Evidence for Observation \\ of Virtual Radio Cherenkov Fields}

% use optional labels to link authors explicitly to addresses:
% \author[label1,label2]{}
% \address[label1]{}
% \address[label2]{}

 \author{Alice Bean, John P. Ralston and James Snow}

 \address{Department of Physics and Astronomy, University of Kansas, Lawrence KS 66045}%Lines break automatically or can be forced with \\
\begin{abstract}

We present evidence for observation of virtual electromagnetic fields in the radio domain from experiment T926 at the Fermilab Meson Test Beam Facility.  Relativistic protons with 120 GeV energy traversed a sealed electromagnetic cavity and were observed in the radio regime of 200MHz-GHz.  Closely related to ordinary Cherenkov radiation, which we also measured, the virtual fields require no acceleration for their existence.  The experiment is also the first observation of fields from hadronic showers, an independent and new confirmation of coherent radio emission from ultra-relativistic particles.   Conditions of very low signal to noise were overcome by a novel and unbiased filtering strategy that exploits exhaustive studies of correlations in the noise backgrounds. Linear scaling of the signal region with the number of beam particles provides evidence of coherence. Extrapolation to measurement of the field of a single relativistic proton charge
is consistent within errors. Our study also illustrates new data processing methods that may be applied broadly in conditions of extremely low signal to noise.

\end{abstract} 

\end{frontmatter}

% main text

\section{Introduction}

We conducted experiment T926 at the Fermilab testbeam facility to measure electromagnetic fields of relativistic bunches of protons passing close to radio antennas.  The fields of moving charges were studied both in free space (virtual fields) as well as with radio antennas embedded in wax (Cherenkov fields).  We present evidence for observation of the virtual electromagnetic fields from the relativistic protons traveling in air. We compare these measurements with detection of Cherenkov radiation, where particles move at about 150$\% $ of light speed in the medium.  Measurements made in the radio frequency regime find the real and virtual signals to be comparable. In particular we find no evidence for dramatic differences between the ``on-shell'' Cherenkov fields and the virtual fields. Besides the conceptual interest of clarifying basic physics, our study also highlights new signal filtering strategies applicable to many circumstances. The signals are exceedingly small; combining standard averaging methods with a novel and highly efficient filtering strategy yields the observation. 

Radio frequency (RF) signals from ultra-relativistic particle showers are the leading technology for detecting ultra-high energy cosmic ray neutrinos. RICE\cite{rice} is the prototype ice-target radio-neutrino telescope.  RICE operates at the South Pole and has established the world's tightest bounds on neutrino fluxes above $10^{18}$ eV.   ANITA\cite{gorham} is a balloon-born instrument with the same purpose. ICECUBE\cite{icecube} is the $km^3$ neutrino detection experiment underway at the South Pole currently developing a radio detection component AURA.  Previous relativistic RF experiments have measured {\it coherent radio Cherenkov radiation} (the ``Askarian effect'' \cite{askarian}) using electron beams with approximately $10^{8}$ times more charged particles than available to us.  We initially proposed\cite{radhep} to use the full Fermilab beam towards calibration of neutrino signals.  Here we describe a much more difficult experiment given the features of the facility available.  To analyze the data we developed signal processing methods that could also ameliorate anthropogenic noise affecting RICE, ANITA, ICECUBE, and numerous other experiments.  The technique amounts to using noise correlations that are not ideally random against noise itself.

\subsubsection{Conceptual Background} 

Cherenkov radiation is a familiar tool of high energy and nuclear physics.  By far the most common use of Cherenkov radiation comes at optical frequencies.  The distance from source to detector is normally many millions of wavelengths, and practically indistinguishable from light of a source at ``infinity.'' The experiment we will describe explores interesting conceptual issues of Cherenkov fields and what is meant in physics by the term ``radiation.''

``Radiation'' is conventionally defined by fields caused by acceleration of charges, that fall like the inverse distance from the source, and that move at the speed of light. There is a certain arbitrariness in these criteria.  Indeed the distinction of virtual
and Cherenkov fields is partly one of terminology.  One defining feature of
"virtual" fields is that they cannot propagate independently to infinity.  The virtual fields we explore are the electric field from moving charges, which therefore need not move at the speed of light.  These fields obviously do not require acceleration of charges for their existence. Virtual fields are commonly associated with ultra-short distances and quantum fluctuations, but their existence is much more general.  The key to measuring virtual fields consists of controlling the environment around the moving charges and working in a low frequency regime where virtual fields extend to macroscopic distances.  Ordinary radio-frequency instrumentation suffices to measure the virtual fields. Despite the simplicity of the situation we have not found a reference for measurement of the impulsive fields from uniformly moving charges under free-space conditions.

The experiment we will describe involves a straightforward measurement of fields moving in free space at about 99.996 \% of light speed, compared to fields moving in wax at about 150\% of light speed in the medium.  Common understanding of Cherenkov physics might lead one to expect qualitatively different behavior, but that understanding hinges on ``real radiation'' propagating to infinity.  Theory suggests the virtual fields should not be very different.  Besides the inherent interest, our experiment sought to measure electromagnetic fields in the radio frequency regime from hadronic showers, and is the first to do so.  Except for scaling due to particle numbers, the virtual fields of hadronic showers are also not expected to be dramatically different from fields or protons in free space. 

The concepts require going beyond the plane wave and dipole approximations of textbook radiation theory.  To develop more general solutions let $A^{\mu}(\vec x, \, t)$ be the vector potential in a Lorentz gauge.  Seek a configuration translating uniformly along the $z$ axis with arbitrary speed $v$: \ba A^{\mu}(\vec x, \, t)= A^{\mu}(\vec x_{T} )e^{i \omega(z-v t)/v}.   \nn \ea    Apply the wave equation to find \ba  \nabla_{T}^{2}A^{\mu}(\vec x_{T} ) +{ \omega^{2} \over c_{m}^{2} }(1-{c^{2} \over v^{2}})A^{\mu}(\vec x_{T} )=0.  \label{transverse} \ea  Here $\nabla_{T}^{2} $ is the Laplacian for transverse coordinates $\vec x_{T}$, and $c_{m}$ the speed of light in a medium.  A plane wave ansatz replaces $\nabla_{T}^{2} \ra -\vec k_{T}^{2}$ giving $\omega^{2} +v^{2}\gamma_{m}^{2}\vec k_{T}^{2}=0$, with $\gamma_{m} =1/\sqrt{1-v^{2}/c_{m}^{2} }$.  An on-shell solution requires $v>c_{m}$, from which trigonometry yields asymptotic plane waves that move away at angle $cos\theta_{c}= c_{m}/v <1$ at speed $c_{m}$. The on-shell case reproduces Cherenkov radiation (actually predicted first by Heaviside) \cite{landau}, where in a medium $c_{m} =c/\sqrt{\epsilon_{\omega} \mu_{\omega}}$, with dielectric constant $\epsilon_{\omega}$.  Such fields satisfy the classic meaning of {\it radiation}. 

Our main focus is {\it subluminal} velocities, $v<c$.  The relevant wave packets are {\it not} reducible to plane waves.  ``On-shell'' solutions with real $\omega,\, \vec k_{T} $ {\it do not exist}, so that the actual solutions are virtual fields.   Cylindrically symmetric solutions to Eq. \ref{transverse} are \ba 
E_{\omega}^{z} ={  -2i q \omega F_{\omega} e^{i \omega z/v }  \over \epsilon v^{2} \gamma^{2}} \ [ K_{0} ( \,  {  \omega |\vec x_{T}| \over v \gamma} \,) - K_{0} ( \,  {  \omega R \over v \gamma} \,)  {  I_{0} ( \,  {  \omega |\vec x_{T}| \over v \gamma} \,) \over I_{0} ( \,  {  \omega R \over v \gamma} \,)  } \ ]  , \nn \\ \:\:  \label{solution} \ea  where $\vec E_{\omega}$ is the electric field in the frequency domain and $F_{\omega}$ is the form factor from longitudinal charge distribution.  Here $K_{0}$ and $I_{0}$ are modified Bessel functions. The singularity of the $K_{0}$ term as $x_{T} \ra 0$ represents a point charge $q$.  Parameter $R$ implements boundary conditions controlling $E_{\omega}^{z}(R)$. The case $R \ra \infty$, an infinite homogeneous medium, has exponentially damped solutions going like $exp(- \omega |\vec x_{T}|/v \gamma_{m})/\sqrt{2 \pi  \omega |\vec x_{T}|/v \gamma_{m}}. $  Otherwise the $I_{0}$ term accounts for the interior solution for cylindrically conducting walls that enforce the boundary conditions, also known as image charge effects. We arranged experimental geometry so that the $I(0)$ term is negligible, and the fields are essentially those of free space.  At the same time we used a conducting cavity to shield external noise. 

In the impulse approximation we measure these fields of uniform motion.   From causality the tiny effects of energy conservation must be accounted for {\it after} the event, during a time set by the inverse frequencies involved.  The transverse extent of the virtual fields $\Delta x_{T}$, normally assumed microscopic, may extend sideways into the macroscopic domain.  Inspecting Eq. \ref{solution}: \ba  \Delta x_{T} \sim 3 \times 10^{-4} cm \, {v \over c}{   \gamma_{m} \over 100 }{ eV \,  \over \hbar  \omega} . \label{range} \ea  Choosing $\omega \sim GHz$ in the radio frequency domain and $\gamma>>1$ makes $\Delta x_{T}$ macroscopic in reach.  Given the frequency dependence of Eq. \ref{range}, radio antennas become the detection tool of choice.  Notice that this radiation is present {\it en vacuo} and (if technology permitted) detectable ``at walking speed''.

\subsubsection{Air Versus Wax} 

A brief examination of Eqs. \ref{transverse} and \ref{solution} shows that the medium enters in terms of the dielectric constant $\epsilon$ and the effective boost parameter $\gamma_{m}$.  Its definition is  \ba   \gamma_{m} ={1 \over \sqrt{1-v^{2}/c_{m}^{2} }}. \nn \ea Charges moving faster than light speed in the medium have $v>c_m$ and $\gamma_{m} \ra i \, |\gamma_{m}|$ becoming imaginary, assuming $\epsilon$ is approximately real. This transformation is exactly like the ÒforbiddenÓ case of imaginary $\gamma$ often said to be absurd in free space, and cited as prohibiting particles with $v>c$. The appearance of imaginary $\gamma_m$ is not absurd when used in the field expressions, and the imaginary phase simply represents energy loss to the medium. 

In wax at radio frequencies the refractive index is about 1.5.  The value of $\gamma_m$ is a frequency-dependent complex function with magnitude of order unity.   The near-field numerical repercussions of $v>c_{m}$ (Cherenkov fields in wax) versus $v \lesssim c$ (virtual fields in air) turns out to be rather modest, a change of ``relative order one.'' 

We found it interesting and somewhat counter-intuitive for short distance Cherenkov fields in wax and virtual fields in air {\it not to be extremely different}.  A simple physical picture explains the physics.  Cherenkov radiation is invariably pictured as a ``shock-wave'' seen asymptotically far from the source. In that region waves constructively interfere along the Cherenkov cone.  This is the regime where photons evolve to configurations called ``real'', with each frequency $\omega$ eventually moving at speed $c_{m}(\omega)$ in directions along the normal to the expanding cone. Eventually the conditions of ``radiation'' are fulfilled, including energy transport away from the center and a $1/r$ scaling of amplitude with distance $r$. 

In the zone close to the detectors, however, there is no shock wave. The phase relations of different frequencies order the fields so that those in air and wax are not very much different.  An analogy with the wake of a moving boat is quite accurate.  The wake is attached to the boat, and moves longitudinally with the boat at whatever speed the boat moves. The virtual field wake for charges moving at 99.996 \% $c$ in air is just the boosted Coulomb field, a kinematic consequence of motion.  An ideal, lossless wake reconstructs itself coherently and does not transport any energy to infinity.  Compare super-luminal motion of charges in wax. The Cherenkov field {\it continues to move longitudinally at the basic charge's speed of 99.996 \% $c$}, and near the charges is comparable in amplitude to the case of air.   However the field lines are drawn back by the source outstripping the propagation speed, setting themselves into shape later to add coherently on the cone, long after the charge has passed.  Just as an airplane does not hit a brick-wall when its speed becomes supersonic\footnote{Supersonic shock wave modeling for airplanes involves a non-linear component, but the primary effect is linear just as in electrodynamics.}, but makes a big ``boom'' far away, the transition from virtual to real radiation close to the source is undramatic.  

It was unnecessary, and we did not attempt to discriminate between the beam speed of 99.996 \% $c$ and $c=1$, a task well beyond our timing resolution.  The experiment we will describe was also not intended to develop fine resolution of overall normalization constants, and our expectations included seeing Cherenkov signals and virtual signals of about the same magnitude. Normalizations are difficult because the simulation (described below) folds together a number of factors that include amplifier gains, antenna response, and beam-antenna separation.  We planned an experiment where it would be sufficient to see the fields ``attached'' to the proton beams to verify detection.

\subsection{Experimental Overview}

\begin{figure}

 \includegraphics[width=5in, height=4in]{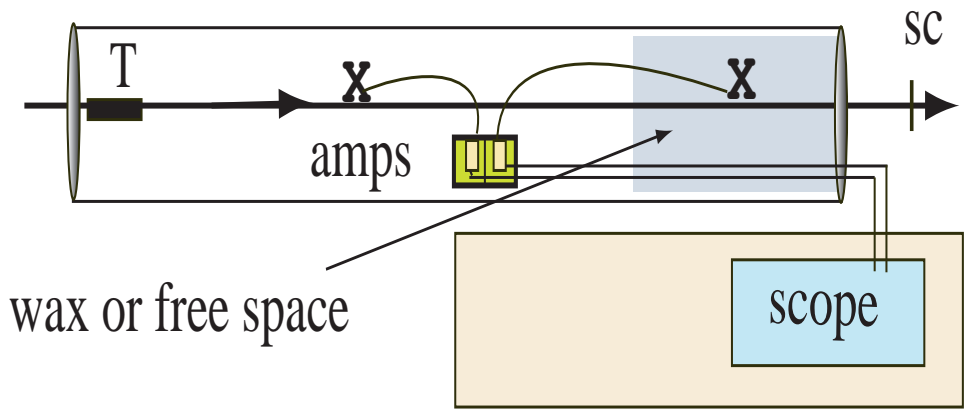}

\caption{\small The tank apparatus in schematic.  The beam enters on axis of a conducting cylindrical cavity 15 feet long containing two antennas ($X$) near the axis.  The trigger scintillator (sc) is 15 feet downstream.  A ``medium'' of free space (``air'') or wax fills the back 3 feet of the tank; a lead pre-shower target (T) was available in the front.  }
\label{fig:Apparatus}
\end{figure}

To measure the fields associated with moving charges we constructed a specialized apparatus (``tank'', Fig. \ref{fig:Apparatus}) at the recently established Fermilab Meson Test Beam facility\cite{testbeam}.  The facility culls 120 GeV protons from the Main Injector and 
steers them towards the fixed target line.  Beam intensity is limited using a collimator, followed by 300m of free flight outside a beampipe to the testing area.  About once a minute, several Main Injector turns (typically 8-11) are built up and extracted in ``fastspill'' (maximum intensity) mode of 7-11 RF buckets, separated by 18.9 ns, a 53 MHz repetition rate.  The first bucket contained $N_{p} \sim $ 300-600 protons, with subsequent buckets being comparable and varying slightly from spill to spill.  Our experiment was the first run at the facility.

The detected signal in volts $V(t)$ is the convolution $$V(t) =\int d\omega \,  \vec E_{\omega}\cdot \vec T_{\omega}e^{-i\omega t}. $$ Here $\vec T_{\omega}$ is the vector valued, frequency-dependent transfer function from antennas, cables, amplifiers, and filters.  From Eq. \ref{solution} the time scale of passage of a 100 GeV proton at 1 cm distance is of order $10^{-11}s$. The Fourier domain pulse is basically flat up to 1 THz, appearing to be a delta function spike per proton. The pulses add coherently, convolved with the time-structure of the beam, discussed shortly. 

The tank apparatus (Fig. \ref{fig:Apparatus}) enforces boundary conditions of cylindrical symmetry about the beam trajectory.  Early plans\cite{radhep} called for a large open volume to measure the transverse field behavior.  To exclude a significant RF background at 53 MHz we adopted a thick-walled aluminum tube of diameter 48 cm, putting the 53 MHz noise well below the minimum (cutoff) oscillation mode of 154 MHz. Two 1 inch long brass biconical antennas are oriented transverse to the beam and positioned about 1 inch from the centerline, separated longitudinally by 10 feet, for redundancy and the possibility of measuring causally correlated signals.   A 10 foot long quad shielded coax cable connected each antenna to its own 200 MHz high pass filter and 50dB amplifier contained in a shielded box within the tank.  Another 40 foot long shielded cable brought each amplified signal to our TEK7104 1GHz analog bandwidth oscilloscope in the test beam control room. Data acquisition was triggered by a 2''x2'' scintillator centered on the beam and positioned 15 feet {\it downstream} of the tank.  Thus RF noise from the phototube could not appear in the antennas until after the beam passed.  The trigger RF spectrum was also measured to be in the regime below 200 MHz. Our data recorded simultaneous readout of 3 channels (2 antennas and the trigger output) with 2500 time points sampled at $\Delta t$=0.16ns intervals for a total event record of 400ns.  Cable delays were chosen to put the scintillator trigger time near the expected first antenna hit and near the center of the event record.  This placement retained an event-by-event ``noise region'' in the first part of each record, well separated by causality from the ``signal region''.  Extensive collection of noise was a key element of the experiment.  

\begin{figure}
\begin{center}
 \includegraphics[width=5in,height=2.5in]{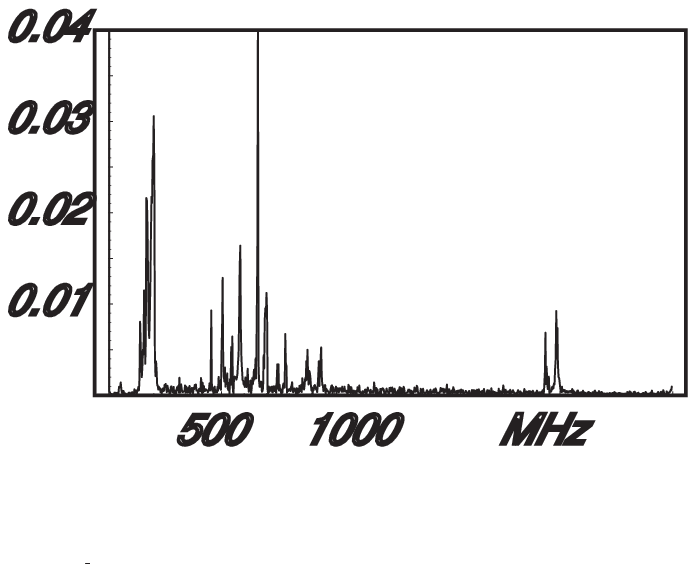}

\caption{ \small Fourier absolute spectrum (absolute value of Fourier transform) of the average of 400 noise segments in channel 1 in arbitrary units.  The peaks above 420 MHz are tank resonances.  The structure around 200 MHz is a mode below the nominal cutoff frequency.}
\label{fig:FourierFNAL}
\end{center}
\end{figure}

\subsubsection{RF Studies} 

Despite heavy shielding, noise of manmade origin ({\it e.g.} cell phone bands) dominated.  We installed metal endcaps on both ends of the tank.  A 1 inch diameter hole in the front cap was covered with 0.040'' thick aluminum to shield radiation while minimizing beam interactions.  Calculations solved the boundary-value problem for our geometry to see if transition radiation would be an issue.   Unlike the usual case for optical frequencies in dielectrics, transition radiation is negligible in our circumstances.  Peaks in the frequency spectrum calculated for the cavity were found to be in good agreement with lines observed in the noise (Fig. \ref{fig:FourierFNAL}). In particular the ``fundamental'' modes dominate the central region.  Identifying every single line is difficult since each mode of given transverse quantum numbers has numerous submodes of longitudinal oscillations.  A discrepancy with expectations occurs in the region of about 200 MHz and below the nominal tank cutoff frequency.  The explanation is leakage of exponentially damped modes (complex longitudinal wave number $k_{z}$). Sample to sample variations in the amplitudes were large. They encode background conditions of myriad origins, which are {\it not} described by textbook assumptions of ``uncorrelated'' noise. 

\begin{figure}
\begin{center}
 \includegraphics[width=5in]{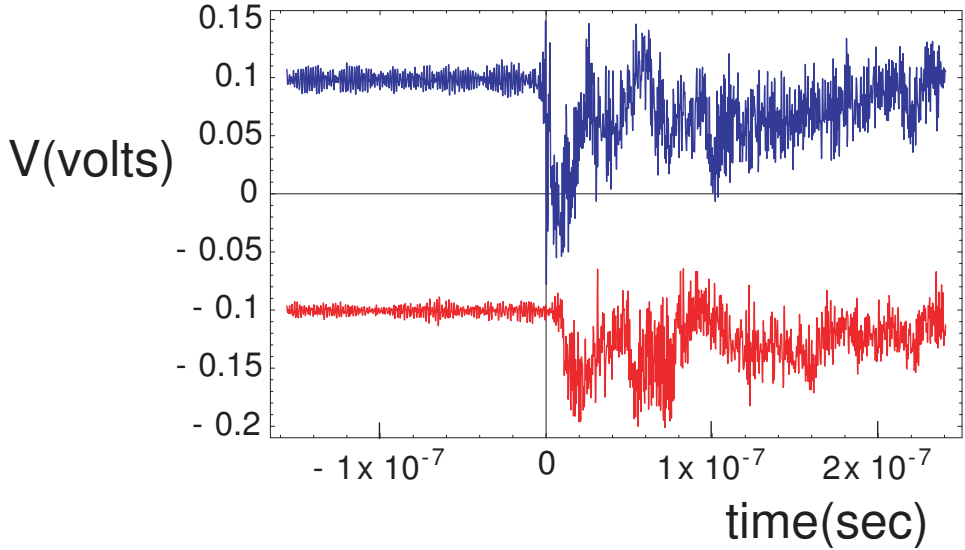}

\caption{ \small Average antenna response over 138 runs of ``click files'' made with a piezoelectric spark to calibrate the system. Top and bottom are channels 1, 2 (upstream, downstream) with timing offset showing light-speed separation of the antennas. The vertical black line shows the trigger location. Units on the horizontal axis are seconds, vertical axis volts, in arbitrary scale. }
\label{fig:Clicks-clfch1av.eps}
\end{center}
\end{figure}

We pre-calibrated the system's transfer function in the lab with a pulse generator.  This left uncertainties of order a few dB from the coupling to the modes in the tank and its local environment.  System response, denoted $T_{\omega}^{click}$, was remeasured directly from the ``click'' spark of a piezoelectric cigarette lighter, which is close to a delta-function in time, calibrating the tank-antenna-amplifier-filter-cable assembly in one step.  The time structure of response is directly shown to be much shorter than the beam bucket structure. Fig. \ref{fig:Clicks-clfch1av.eps} shows the average of 138 click-runs. 
 In these and other figures the vertical scale (units of volts) depends on amplifier gains and cable attenuation and so is plotted in arbitrary scale.  We naturally use the same scale in passages about quantitative comparison. 

The ``click runs'' also verified the expected time difference of a causal signal between the two antenna channels, approximately 10ns (Fig. \ref{fig:Clicks-clfch1av.eps}).  Thus our calibration, and subsequent data processing, are based on truly radiative processes. 

\subsubsection{Target, Beam and Background Noise}

The test-beam cycle of one spill per minute and a fixed total running time dominated our total number of events.  We collected two datasets of approximately 400 events, treated independently in the analysis. {\it Set-A} used antennas in an empty tank (``air'') to look for virtual radiation.   In {\it Set-W}  the downstream antenna was embedded in paraffin wax extending 3 feet in front of the downstream antenna.  Wax is a radio-transparent medium with index 1.5 at GHz frequencies, producing radio Cherenkov radiation with relativistic proton beams.  In Set-W a 16 cm long lead target was also placed at the beginning of the tank to generate a hadronic shower. The target length was calculated to produce about one nuclear interaction on average.  Yet some events will have more than one interaction, and previous studies \cite{razzaque} have shown that radio frequency radiation from fully developed hadronic showers eventually imitate electromagnetic ones.  The pre-radiator was designed to produce a modest signal enhancement of order one on average, with occasional fluctuations that we hoped might stand out when data was processed.

\begin{figure}
\begin{center}
 \includegraphics[width=5in,height=2in]{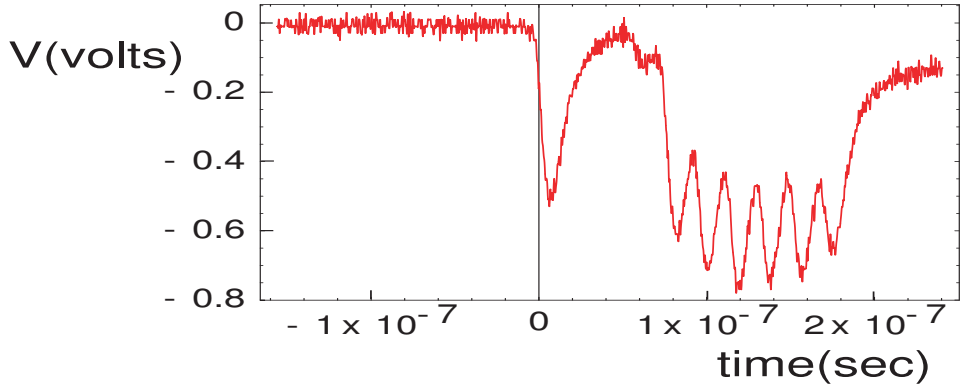}

\caption{ \small The bucket structure determined from a typical phototube response.  The average response over hundreds of runs is practically indistinguishable.  Units on the horizontal axis are seconds, the vertical axis in volts.   }
\label{fig:Ch3.eps}
\end{center}
\end{figure}

However the central spot of the beam traversed the tank axis with fluctuations of order 1 inch, as surveyed directly with scintillator finger counters.  Beam variations themselves created an order of magnitude uncertainty in the normalization of the expected signal.   A plot of typical phototube response showing the bucket structure is shown in Fig. \ref {fig:Ch3.eps}. We do not use the phototube response and click runs {\it per se} in our data processing.  They are building blocks for a simulation made separately that serves as an order of magnitude consistency check which happened to work out quite satisfactorily.

\begin{figure}[htbp]
\begin{center}
\includegraphics[width=4in,height=4in]{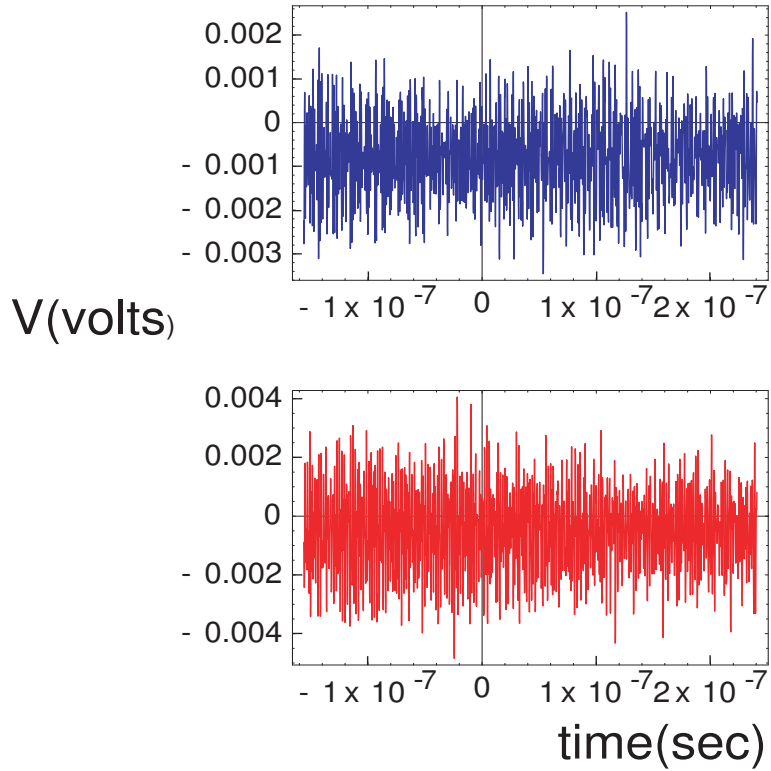}

\caption{\small  The average voltages of channel 1 (top) and channel 2 (bottom) for 361 good runs in air. There is no visible trace of a signal beginning near the center of the time window. Units on the horizontal axis are seconds, the vertical axis in volts.  The vertical black line shows the trigger location.  }
\label{fig:Bodch1Ch2Av.eps}
\end{center}
\end{figure}

The minimum rms noise values of around 30mV, compared to signal level expectations of order 100 $\mu V$, pushed the experiment into a regime of very low signal to noise ($S/N$).  Visual inspection of the trigger traces rejected events with grossly fluctuating or incomplete buckets.  The remaining events were averaged to construct $\bar {V} (t) $ in which the rms noise was reduced to approximately 1.5mV.   Fig. \ref{fig:Bodch1Ch2Av.eps} shows the averages in the two antennas of 361 good runs in air. Even after averaging there is no visible trace of a signal beginning near the center of the time window.

\subsubsection{Observations During the Run}

Careful visual observation played a role during running.  Good, well centered events were accompanied by a barely discernable 53 MHz structure we called ``picket fences.'' 
Picket fences mysteriously survived explicit software filtering which deleted the 53 MHz region, and were eventually identified as {\it beat frequencies}, due to the product of the bucket form factor's 18.9 ns repetitions and the transfer function.  By picket fence observation we could center the beam to about 1 inch accuracy, an empirical detection without need for signal processing.  When later classifying events using offline software, picket fence behavior was also found in 30-100 noise files, wherein pickets extended across the whole range of the data, including the pre-causal onset region. Subsequent consultation with the Fermilab staff\cite{staff} confirmed the likely existence of a precursor associated with the formation of buckets and occurring whether or not we received the beam.  Removal of these files made no difference to signal detection in the causal onset region except for dilution of statistics.  They also cannot be an explanation of our observations in the signal region because the pickets extended over the entire data record, noise-region and signal region alike. Beam centering done with the beam-on picket-fences actually made an unexpected consistency check. In these runs a hypothetical phototube signal would remain the same, as the trigger scintillator paddle was bigger than the beam displacement.  

\subsection{Results Overview } 
 \label{sec:charge}
 
In Section \ref{sec:Analysis} we will describe the analysis strategy we devised to extract a signal under high noise conditions.  To simplify the presentation we briefly summarize some results. 

 \begin{figure}[htb]
\begin{center}
\includegraphics[width=4in]{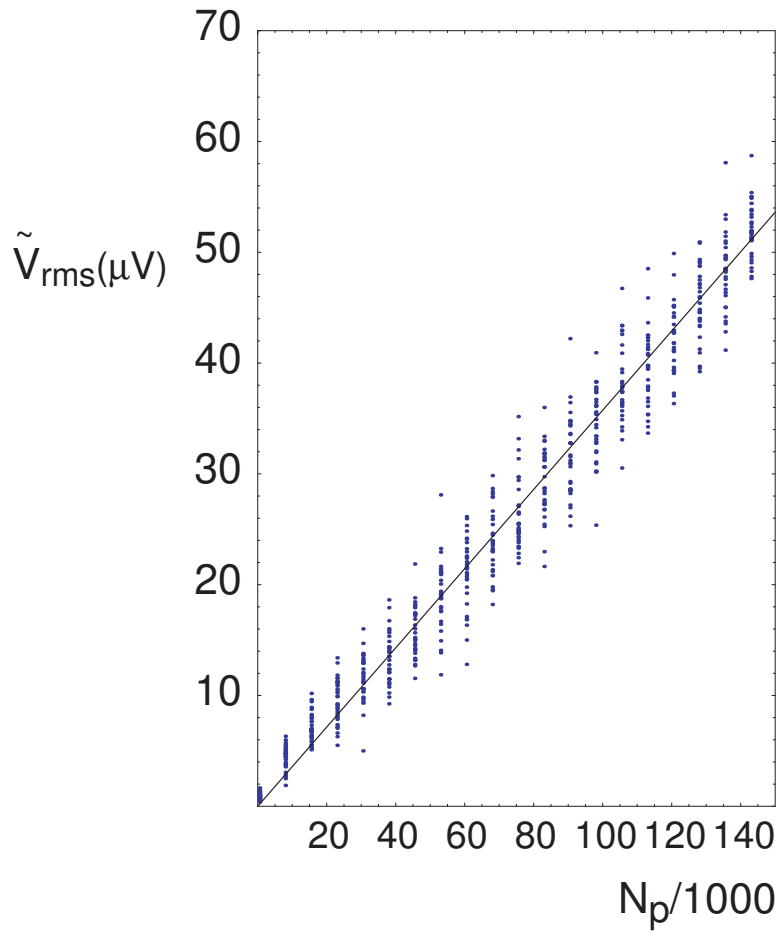}
\caption{ \small Linear scaling in the signal region. Voltage versus particle number $N_{p}$, in thousands of protons, normalized to 500 protons per run. The square-root random-walk contribution has been subtracted. The slope of the straight line fit has been predicted by the simulation within a factor of 2.  Air Ch 1. }
\label{fig:LinearScaling.eps}
\end{center}
\end{figure}

Figure \ref{fig:LinearScaling.eps} shows the filtered $rms$ voltage $\tilde V_{rms}$ extracted from the signal region of the data, as predicted by causal arrival of the fields moving with the beam. The figure shows $\tilde V_{rms}$ as a function of the number of protons $N_{p}$ used in the analysis.  The linear dependence of $\tilde V_{rms}$ on $N_{p}$ is evidence of {\it coherence}, by which the voltage measured is proportional to the total charge producing the field. The slope of the plot is directly proportional to the number of particles per run, the amplifier gains, antenna response, and finally the elementary proton charge $e$. 

\begin{figure}
\begin{center}
\includegraphics[width=4in,height=4in]{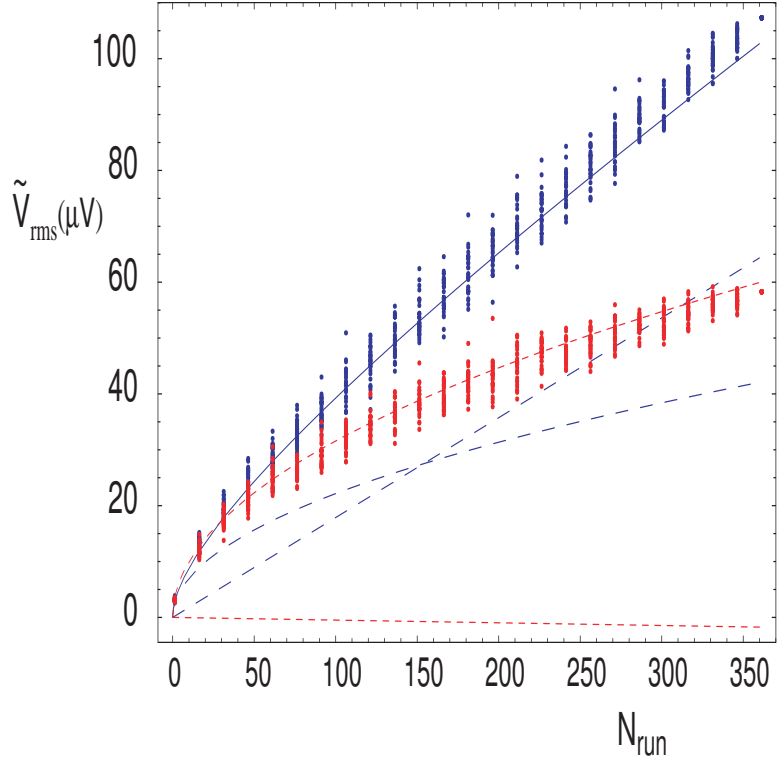}
\caption{ \small Dependence of $rms$ filtered voltage $\tilde V_{rms}$ on the number of particles, as represented by the number of randomly-permuted runs $N_{runs}$ analyzed. Linear coherent scaling with $N_{runs}$ of the signal region (top curve, blue online) is evident. Scaling with $N^{1/2}_{runs}$ occurs for the noise region (bottom curve, red online).  The separate contributions of $\a N_{runs}^{1/2}$ and $\b N_{runs}$ are shown for comparison (dashed lines).  Air Ch 1. }
\label{fig:Bod1Coherence.eps}
\end{center}
\end{figure}

In making this figure a term scaling like $N_{p}^{1/2}$ and consistent with the noise contribution has been subtracted. Fig. \ref{fig:Bod1Coherence.eps}
illustrates the separation into noise and signal components using $N_{runs} =N_{p}/500$.  Noise contributions are consistent whether they are obtained from a fit to the signal region or whether obtained from the pre-onset parts of the data (``noise region'').  The quantity $\tilde V_{rms}$ itself has been developed by signal/noise improvement procedure (Section \ref{sec:method}) that sorts and then rejects dominant timing patterns found among correlations of the noise region.  Finally (Section \ref{sec:timestucture}) the timing pattern of the signal region is consistent with the pattern predicted by the simulation, using a $\chi^{2}$ test.

The overall normalization of our simulation is rather uncertain, and deliberately does not involve a very refined procedure.  Factors contributing to the uncertainty are the beam-antenna separation, the number of protons per bucket, and the antenna-amplifier gains.  Several factors multiply and it is difficult to control the normalization to better than a factor of 10 or so.  Nevertheless we went through the exercise and committed to a value before examining the data.  As Fig. \ref{fig:ChiSquareBod1.eps} in Section \ref{sec:shape} shows,  the value of the slope of Fig.\ref {fig:LinearScaling.eps} has been predicted by the simulation up to one re-normalization factor $n \sim 2$.  

All four channels show behavior very similar to the one illustrated in Fig. \ref{fig:LinearScaling.eps}, \ref{fig:Bod1Coherence.eps} highlighting Air Ch 1.  We were not surprised to see similar linear scaling in both Air and Wax, but a word of explanation might be in order.  As emphasized earlier, the phenomenon of Cherekov radiation and the impulsive fields of virtual free space radiation are not fundamentally different in the near zone.  Linear coherence is expected in all cases because the total fields are proportional to the total charge. Only at distances of ``far zone'' and beyond does the small virtuality of fields in Air lead to their confinement, or to propagation out to infinity in Wax. 

It is interesting that by extrapolation to one particle the experiment can then measure the charge of single relativistic protons.  The extrapolation may seem bold, bit it is justified once we have established linear behavior on the number of particles.  Our experiment is like Millikan's with larger relative errors. The data analysis reached its limit when the statistical errors are smaller than the irreducible systematic errors.  Using the normalization of the simulation predicted, and adding a generous error of a factor of 2 on the predicted slope, the experiment yields an overall measurement of the single proton's charge of order $(2\pm 8)\times 10^{-19} $ C.  
\section{Analysis}
\label{sec:Analysis}
Given low $S/N$ it was necessary to devise new strategies to resolve a signal. 
In the following we provide the conceptual background.  This will be followed by description in depth of a novel data filtering strategy: \begin{itemize} 

\item We sought a method not intrinsically tied to Monte Carlo simulations. The role of the Monte Carlo is to provide a secondary consistency check, and not to be the primary basis of declaring a signal. 

\item We realized that extensive ``noise files'' taken during the run should be used to define a signal self-consistently and in a mode of data-versus-data, not data versus theory. 

\item We sought a method that would not be overly sensitive to small variations of cuts, parameter choices, or subjective judgments, while still creating latitude for judgment to be used.

\item We desired a method based on a  ``blind'' procedure, treating signal data and noise under one uniform method without using the signal data to define a signal. 

\end{itemize}

\subsection{Method Overview}
\label{sec:method}

The criteria are carried out by methods based on ``filters,'' which broadly mean algorithms discriminating on the basis of patterns in the data sequence.  

A huge literature exists on ``matched filters.'' Matched filters preferentially pass patterns of a predetermined class. Analog radio receivers essentially operate with Fourier-component matched filters.  They are successful under very low $S/N$ when signal lies in narrow frequency components, to which the radio oscillator responds resonantly. 

Yet under conditions of low signal to noise, the entire strategy of ``pattern acceptance veto'' is unstable.  By definition random noise generates all possible patterns.  When noise fluctuations that happen to reproduce selected pattern overwhelm the signal,  the strategy will fail. For our experiment the impulsive signal has a broad Fourier spectrum, and is also difficult to model with high accuracy.  The uncertainties from beam-location, bucket fluctuations, and system calibration make a Monte Carlo-based matched filter ``tuned to signal'' unwise.  

We found a new strategy realizing that our noise (and the noise of most experiments) is far from thermal and contains myriad correlations.  Noise correlations can be put to advantage. We devised a method to systematically create ``noise filters '' well-matched to the patterns in noise.  This requires more than one filter, but by exploiting linear combinations of patterns, even a few noise filters can capture an infinite number of noise configurations.  After classifying the noise we created a noise rejection veto 100 \% efficient against the patterns found.  No attempt was made to select a signal pattern: instead, those components untypical of noise were simply left untouched. While some signal will be rejected, the signal/noise ratio can be substantially improved by rejecting relatively more noise than signal.

\subsection{Constructing Filters Using Noise Against Noise}

First define a few terms:  
\begin{itemize} 
\item  A one-dimensional data list of length $D$ is considered a vector on $D$ dimensions.

\item   A {\it filter} is an operation on the list to return a new list.  We only consider linear operations.  Any linear operation on the list can be represented as multiplication by a $D \times D$ matrix.

\item   A ``pattern'' is a vector orthogonal to other patterns, and thereby independent. Orthogonality is defined using the dot product $\bracket{A}{B}=\sum_{j} \, A_{j}B_{j}$. 

\item  On $D$ dimensions there are at most $D$ patterns, which can be used to make an orthonormal basis.   

\item The set of all linear combinations of a subset of patterns, or basis vectors, is called a subspace.  

\item Filters can accept, reject, or manipulate patterns and thereby control subspaces in which data occurs.  

\end{itemize} 

\subsubsection{Extracting Pattern Subspaces} 

Consider summing many instances (label $J$) of a data vector $d_{i}^{J}$ that is ``trying to repeat'' a single pattern $p_{i}$.  If patterns occur with random coefficients $\a^{J}$, then \ba d_{i}^{J} =\a_{J}p_{i},  \nn \ea  and the sum tends to zero.  If the sum happens to be non-zero it can be recorded.  This still leaves all the patterns that cancel undetermined. 

The solution to discovering general data patterns is to make {\it outer} products, namely matrices, and average them. Take a single data string $d_{i}$, and make the matrix $d_{i}d_{j}$. Suppose another data string happens to be $-d_{i}$; it yields the same matrix $d_{i}d_{j}$.  Regardless of coefficients similar patterns {\it add} upon averaging the outer products. 

Summing outer products produces a density matrix\footnote{The term borrowed from quantum mechanics is exact. A study of quantum theory led to the procedure.} $\rho_{ij}$, \ba \rho_{ij} = \sum_{J} \, \a_{J}^{2} p_{i}p_{j} \ra  constant \times  p_{i}p_{j}. \nn \ea   The underlying pattern $p_{i}$ is then recovered as the eigenvector of $\rho_{ij}$.  

When more patterns with arbitrary coefficients exist, a set of patterns that make an optimal basis is again found from the eigenvectors of $\rho$.  To make an ``optimal'' filter, then: (1) develop the optimal patterns from the density matrix of a sufficiently large sample;  (2) classify pattern importance quantitatively; (3) expand the data in the basis of patterns (4) throw away components not desired (normally small, or rare elements); (5) revert the expansion back into the original components.  

\paragraph{White Noise:}  

Textbook ``white noise'', often called simply ``noise'', is defined by correlations of data  $n_{i} $ that has a density matrix \ba <n_{i}n_{j}> = \sigma^{2}\delta_{ij}. \nn \ea The statistical average $< \, >$ is probed experimentally by adding up noise samples $n_{i}^{J}$.  If physical noise would actually satisfy the textbook criteria, then no particular pattern is favored.  All possible patterns are produced equally.  The attempt to extract a special eigenvector of $\delta_{ij}$ yields no particular vector because all vectors are eigenvectors of the unit operator.  

Ideal white noise is actually rare.  For this reason one can classify correlations in the noise to find the dominant patterns that actually exist.  When $S/N>>1$ patterns in the data will overlap with signal.  Then retaining the dominant patterns improves $S/N$.  We faced a situation of $S/N<<1$ that naturally suggested the reverse procedure, and also necessitated optimizing our filters.

\subsubsection{Mathematical Description}  

Our filters use a variation of the Karhunen-Loeve($KL$)\cite{KL,chambers} method from image processing.  Our reverse $KLJ$ filtering is done in three steps: (1) Data from the noise region of the files (the first 900 points) are partitioned into non-overlapping segments of length $D$ (``bins'') containing vectors $\ket{noise^{J}}$.  (2) We construct orthogonal projectors $\pi^{a}$ such that  \ba \,{ < \: \EV{noise}{\pi^{a}}{noise} \: > \over <\, \bracket{noise}{noise} \,>}  = n(a) \ra max,  \nn \ea   where $< \: \: > $ denotes the sum over bins.  
Each projector $\pi^{a}$ defines a one dimensional ``noise subspace.''

Optimization is done by solving an eigenvalue equation\cite{KL}.  Construct the noise density matrix \ba \rho_{noise}=\sum_{J} \, \ket{noise^{J}}\bra{noise^{J}}. \nn \ea   Solve the eigenvalue equation \ba \rho_{noise}\ket{e^{a}} = n^{a}\ket{e^{a}} . \nn \ea    By the Rayleigh-Ritz variational theorem, the eigenvectors maximize $\EV{e}{\rho_{noise}}{e}$, for any possible normalized $\ket{e}$.  We will call the normalized eigenvectors ``noise states'' for simplicity.  Assigning $\pi^{a}=\ket{e^{a}}\bra{e^{a}}$ then gives \ba \EV{e}{\rho_{noise}}{e}=tr(\pi^{a} \rho_{noise})=n^{a}, \nn \ea   where $tr$ is the trace.  The optimal noise subspaces so constructed have the maximum overlap with noise state-by-state.  

The idea of optimal filtering itself is not new.  The process of {\it Wiener filtering} is invariably described as optimal. It is easy to show that when noise is ``ideal'', featureless, and time-translationally invariant, then the noise states will be Fourier modes.  For our purposes it is unwise, inefficient and unnecessary to make such idealized assumptions when one has actual data for the noise. 

We label the noise states by their eigenvalues, sorted from largest noise power to smallest, $n(1)>n(2)>...n(D)$.  We then construct $\pi_{noise}({\cal R})= \sum_{a}^{{\cal R}} \,\pi^{a}$, which is the most efficient noise-passing filter of given rank ${\cal R} \leq D$. (3)  Our filter consists of applying $\pi_{signal}({\cal R})= 1-\pi_{noise}({\cal R})$ to the data, removing ${\cal R}$ dimensions of noise and passing $D-{\cal R}$ dimensions populated by noise as little as possible. In symbols, \ba \ket{V}_{filtered} =  (1-  \sum_{ a   } \ket{e^{a}}\bra{e^{a}})\ket{V} . \nn \ea We scrupulously arranged that the filter construction never uses the signal region of the data.  Under the hypothesis that the entire data set is noise, the filter will not favor the causal onset region over any other.

\subsubsection{Indices and Translational Properties} 

Partitioning the data into ``bins'' of length $D$ loses no information, and is simply a relabeling of indices: \ba data_{i} & \ra & data_{k}^{J}; \;\:\:\:\  J   =  int({i \over D}); \;\:\:\:\ k =mod_{D}(i). \nn \ea Here $int$ takes the integer part of its argument, producing the bin-index $J$.  The function $mod_{D}$ is the remainder of division by $D$, yielding the index within the bin $k$. The inverse of the transformation assigns index $i$ by the rule \ba i = J D+k .\nn \ea In this way filtered bins are conjoined to re-make data in its original index notation. 

Generally the noise states make a complete set for any data in each bin or length $D$.  \footnote{An exception occurs if fewer vectors than the length $D$ are used to make noise states, or if those vectors for some reason lack linearly independent components. This is readily cured by adding more vectors from the noise region.} Thus repeated application of the complete noise state basis bin-by-bin is perfectly lossless. When a filter is applied one deliberately rejects information {\it within bins}, without changing the bin dimension $D$.  To make this less abstract, suppose the filter only retains 2 specific Fourier modes on a $D=10$ dimensional bin.  Suppose the data length is 2000 points. Then on each of 200 bins, those components of the data in each Fourier mode will pass.  Experience processing physical data shows that one overall mean - the zero frequency mode - should be removed, else it overwhelm all other modes. 

After filtering the labels $J, \, k$ can be used, or they can be reverted to the original monotonically running $i$:  \ba data_{i}^{filtered} = \sum_{Jk}\, \delta(i- J \, D+k) \,  data_{k}^{J, \, filtered}. \label{datfil} \ea  At this point the partitioning and binning has done its job and ``disappeared'' leaving revised data in its original format. 

A question arises whether the filtering process depends on the start-point of the bins.  
The answer depends on the nature of the data used to make the filter, and the number of states retained.  The extreme case of retaining one lone state, for example, forces one specific pattern to re-appear in each bin with sign and normalization that is the best possible fit by a very limited subspace. Retaining many states has a strong tendency to be nearly translationally invariant and independent of how binning is started.  One explanation is that special glitches in data between bins, that might distinguish one bin-start point compared to another, tend to be small effects compared to the accumulation of data over the domain of many bins.  The other explanation seems to be that translational symmetry of data correlations is a generically reasonable approximation.  Even so, forcing Fourier modes (Wiener filtering) omits phase correlations that cannot be seen in Fourier power spectra and will degrade performance. 

\subsubsection{Timing Resolution} 

Noise states tend to be ordered in ``smoothness'' and are often similar to Fourier modes.  Filtering removes components needed to make all possible patterns.  Filtering is then guaranteed to downgrade timing resolution within bins. The proof of this is very simple.  Consider the basis with elements that are spikes at each point $k$: $e_{k}^{a} \ra \delta_{a k}$.  Aligning a data element perfectly with such a spike gives perfect timing resolution, namely the sampling time $\Delta t$.  A superposition of many noise-basis elements is generally needed to make spikes, as in Fourier series.  Decimating the basis by filter construction can only downgrade timing resolution compared to examining data spike-by-spike.  Conversely, triggering spike-by-spike on raw data allows the maximum bandwidth of noise to intrude, and will generally produce the lowest possible $S/N$.  Evidently an ``uncertainty principle'' operates which is more general than the usual relationship for the Fourier domain.  

Given the large amount of noise our filters reject, timing resolution inside bins is quite poor, and timing resolution of order $D \Delta t$ is expected.  We made several studies hoping to evade this reality before we realized that the extended time structure of the buckets made it pointless.  We turn to more discussion on how filter parameters were chosen.

\subsubsection{Signal Retention} 

Our method can increase signal to noise when sufficient signal exists in the subspace retained.  One also wants to make sure that tentative signals lifted out of the noise are not overly sensitive to fine details of procedure. These facts determine the useful ranges of bin dimension $D$ and the filter rank ${\cal R}$.

\begin{figure}[t!]
\begin{center}
\includegraphics[width=4in, height=3in]{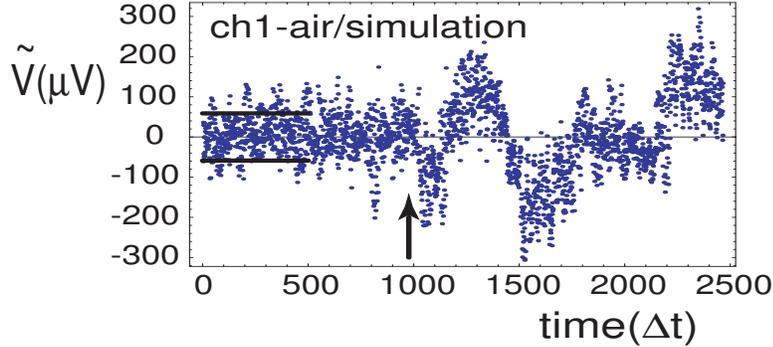}
\caption{ \small Simulation of signal shape post-filtering, channel 1(air).  Earliest signal onset occurs at the arrow (point 970); shape is dominated by the form factor.  Units are $\mu V$;  time in units of $\Delta t$=0.16ns. Horizontal lines show $\pm 1 \sigma$ in the noise region. }
\label{fig:MCfliteredSignal}
\end{center}
\end{figure}

We studied simulated signal events under the actual filters.  Our nominal signal is $V_{\omega}^{signal} = n N_{p} T_{\omega}^{click} \bar F_{\omega}$, where $n$ is a normalization adjustment relative to theory, and 
$\bar F_{\omega}$ is the Fourier transform of the averaged phototube pulse.  In the time domain $\bar F(t)$ is a long bump with one initial and six subsequent bucket peaks repeating at 18.9 ns.  This form factor and the need to average data with timing jitter precluded sharp timing resolution that would simplify analysis. The bin length $D$ was chosen to compromise between noise rejection (improving with larger $D$) and timing resolution (deteriorating with larger $D$).  For $D\lesssim 25$ our signal simulation showed predominant overlap between noise and signal and the filter is too weak.  For $D \gtrsim 28$ separation is good.  Otherwise dependence on $D$ is weak, and we settled on $D=32$. The rank ${\cal R}$ is determined from the signal efficiency $\eta(dB) =10\log_{10}( \sigma_{passed}/\sigma_{raw})$, where $\sigma$ denotes the $rms$ of noise regions passed by the filter compared to raw data.  Set by set, we adjust ${\cal R}$ to be as large as possible to reject noise.  Meanwhile $\eta(dB)$ is required to be rather flat, so that signal efficiency is not overly sensitive to the exact choice of ${\cal R}$.  We insist $ {\cal R}\ra {\cal R} \pm 1$ does not change $\eta( dB) $ by more than about 1 unit.  This procedure fixed ${\cal R}=11$ for Channel 2 in Set-A and ${\cal R}=12$ otherwise.  A simulated signal ($n=3$) plus noise $\tilde V^{signal}$ post-filtering is shown in Fig. \ref{fig:MCfliteredSignal}.  Noise was obtained from pre-onset regions of the data to make this figure.  The filter retains over 95\% of the signal power while reducing the noise by a factor of about 20.  None of this is very sensitive to the details and uncertainty in modeling the signal. It is a generic fact that almost any signal has components different from the details of the physical noise.  To achieve similar results without the noise-rejection filter would require about 400 times more data, the equivalent of running for about 3 years.

\begin{figure}[htb]
\begin{center}

\includegraphics[width=5in,height=5in]{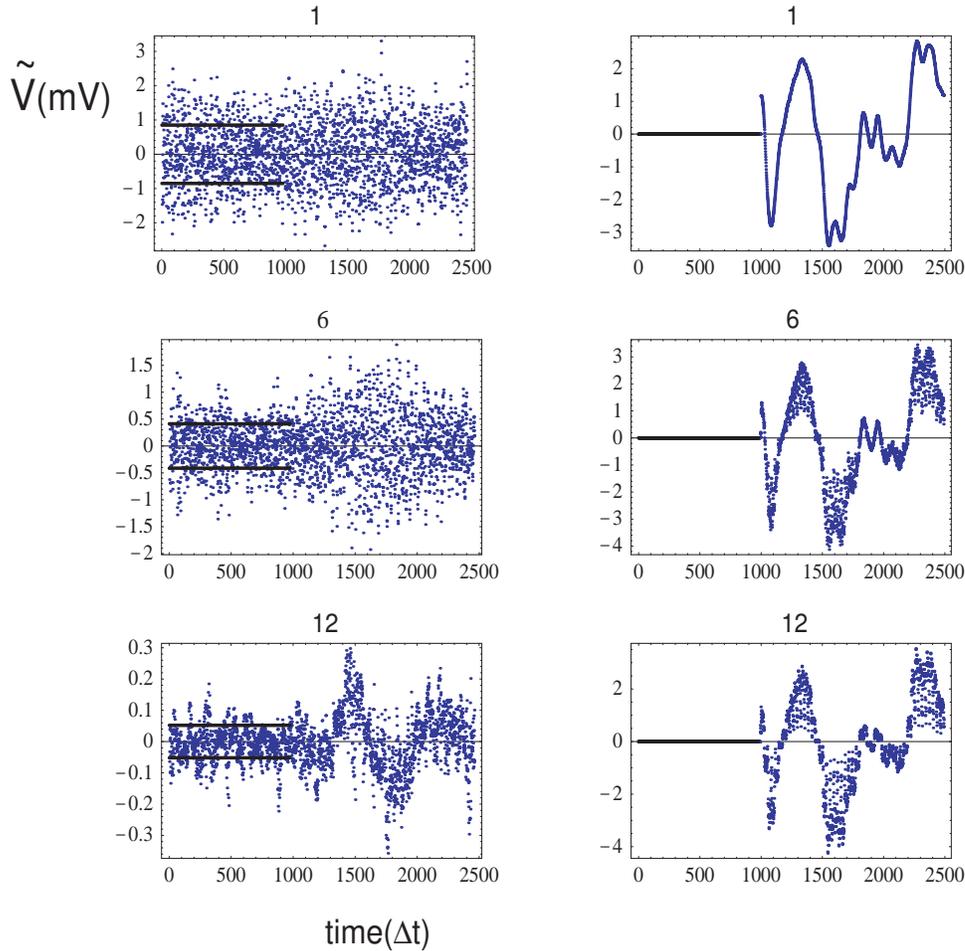}

 \caption{\small  Effects of filter subspaces.  Filters retain noise-state labels equal to or exceeding $N_{cut}$, shown at top of frames. The top rows labeled ``1'' keep all states, the bottom keep $N>12$.  {\it Left column}: Filtered data (air).  Note changes in vertical scale from different filter states.   {\it Right column}:  The filtered form factor, as calculated from a click file and average bucket structure, and using same procedure as the data.  Preferential passage of the signal compared to noise is clear.  Horizontal lines consistently show $\pm 1 \, \sigma$ computed from the first 950 points after filtering.  A constant voltage is removed in filtering, and has not been restored.    }

\label{fig:FilterShow}
\end{center}
\end{figure}

Dependence of the filtering process on noise states retained is shown in Fig. \ref{fig:FilterShow}.  The left column shows data processed by filters that keep noise state labels equal to or exceeding a cutoff $N_{cut}$, shown at top of frames.  Thus $N_{cut}=6$, used for the middle panels in both columns, use filters which omit the first 5 most important noise states. The right column shows the same procedure applied to the form factor, as calculated from click file and average bucket structure.    Since the means are removed before filtering, as discussed earlier, one constant voltage offset shifts the figures. Note changes in the vertical scale of the data, especially compared to the simulated signal.   Preferential passage of the signal compared to noise is clear, increasing the $S/N$ ratio.

\begin{figure}[t!]
\begin{center}

\includegraphics[width=3in,height=5in]{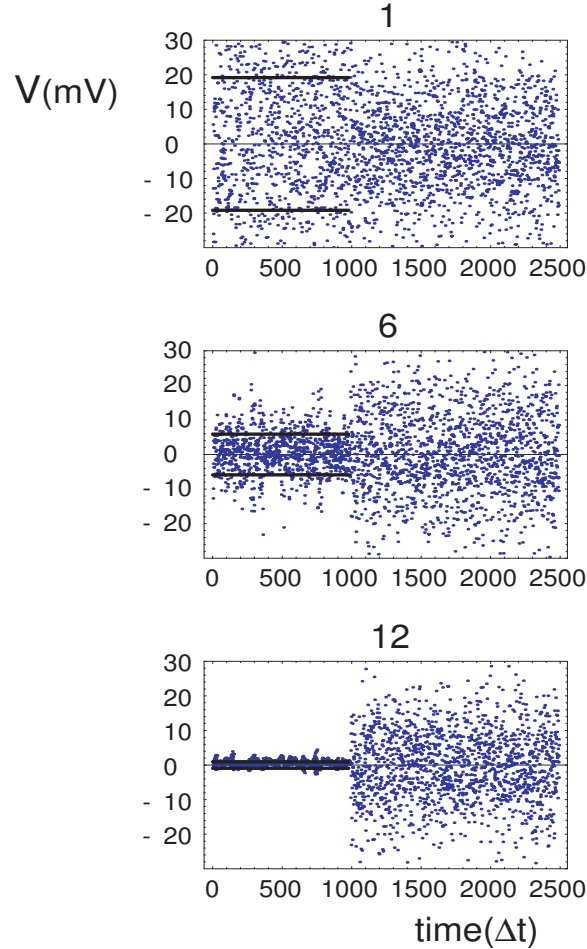}

 \caption{\small  Effects of the filter, as made with different subspaces, acting on the physical data compared to randomly generated numbers.  Labels as in Fig. \ref{fig:FilterShow}. The first 950 points come from a typical data file; the last 1250 points come from a Gaussian distribution (idealized noise) with the same standard deviation.  Physical noise is rejected far more effectively than idealized noise.  Horizontal lines show $\pm 1 \, \sigma$ over the regions indicated.  }

\label{fig:NoiseShow}
\end{center}
\end{figure}

A rather different illustration is given by Fig. \ref{fig:NoiseShow}. The left panels show the action on the physical noise on a typical file. The right panels shows the effect on simulated (textbook) white noise from a Gaussian distribution with the same standard deviation.  Filtering of the physical noise is efficient, while extremely mild on Gaussian random numbers.  As explained earlier {\it perfectly random} numbers fill out all possible patterns and tend to pass any conceivable filter.  Meanwhile {\it physically random} data have correlations that allows far more effective rejection.

\subsubsection{Simultaneous Mode and Time Information}

\begin{figure}[t!]
\begin{center}
\includegraphics[width=5in, height=4in]{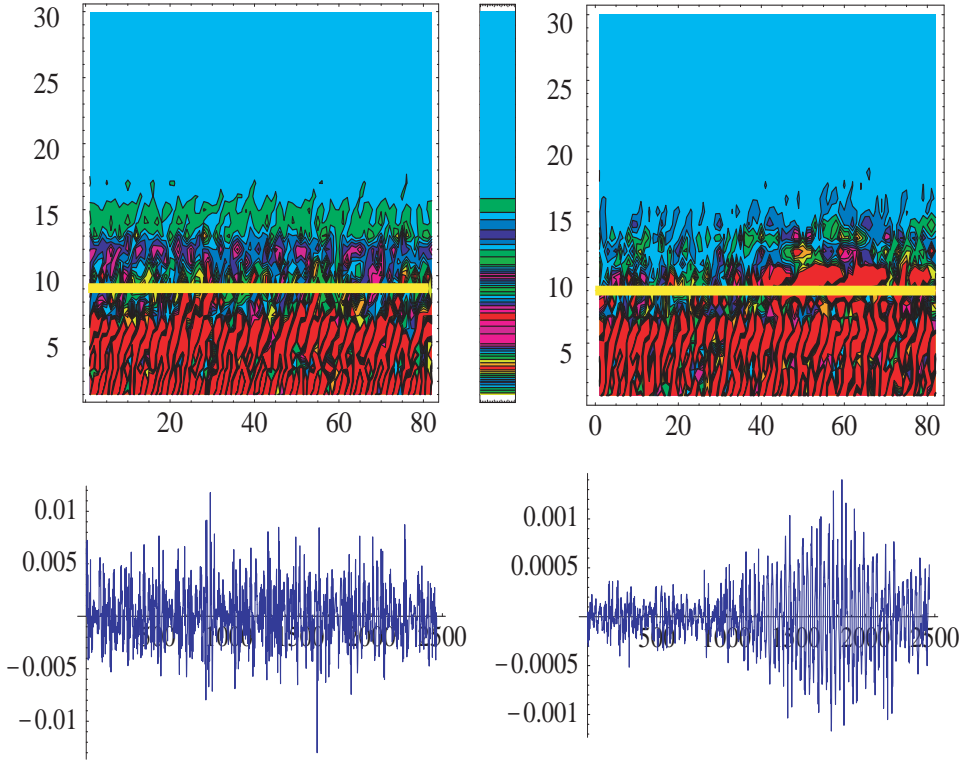}

\caption{\small  Simultaneous mode and time information.  Upper panels show noise-basis projections of the data arranged by rows, from bottom (most noise) to top (least noise).  Binned-time is plotted along the horizontal. Visual appearance suggests keeping components above the horizontal line for {\cal R}$>9$, rejecting integrated power at the 200:1 level, which has been done here to demonstrate insensitivity to fine details of the cut. The thin middle panel shows the single-bin signal click-pattern in the noise basis, expanded horizontally to be visible.  Bottom panels show the data of a filtered typical run (left) and the filtered average over runs (right), ch 1.     }
\label{fig:PlanLayout}
\end{center}
\end{figure}

Our filtering scheme has a rather natural graphical representation.  For each given bin-time, projections of the data into each noise mode ($data_{k}^{J, \, filtered}$, Eq. \ref{datfil}) are plotted on the vertical axis of ``running mode'' plots (Fig. \ref{fig:PlanLayout}).  The noise subspaces are ordered from bottom (maximum noise) to top (least noise).  This is repeated for each bin time plotted along the horizontal axis.  A graphics program then constructs contours of constant amplitude in the mode-versus-time plane.  Rejection of a noise subspace consists of ignoring data below a horizontal line (subspace $\pi_{noise}$).  Detection of a signal within a bin-time or two consists of observing structure in the region above the line (subspace $\pi_{signal}$.)  

We also included a thin middle panel (Fig. \ref{fig:PlanLayout}) showing a single-bin signal click-pattern in the noise basis.  It has been graphically expanded horizontally to be visible. The panel demonstrates that signal protrudes well out of the noise subspace cut by the horizontal line. 

The running mode plot then shows all the data, for all times, organized into noisy and quiet projections.  It is a matter of taste whether or not to square the mode-projections to examine mode power, to plot positive and negative values scaled to make structure visible, to take the logarithm, and so on.  Many options exist.  Fig. \ref{fig:PlanLayout} also makes visible the effects of adjusting the ``cut'' on noise power.  Visual inspection suggests making a cut with {\cal R}$> 9$, weaker than the ones we chose on a quantitative basis.  (This sort of visual inspection is {\it retrospective} towards finding a signal and might have to be justified.)  The figure shows action of the weaker filter is very similar to our analysis,  because we did not choose our cuts in an overly sensitive region.

\subsection{Basic Analysis of Data}

\begin{figure}[!]
\begin{center}
\includegraphics[width=5in, height=4in]{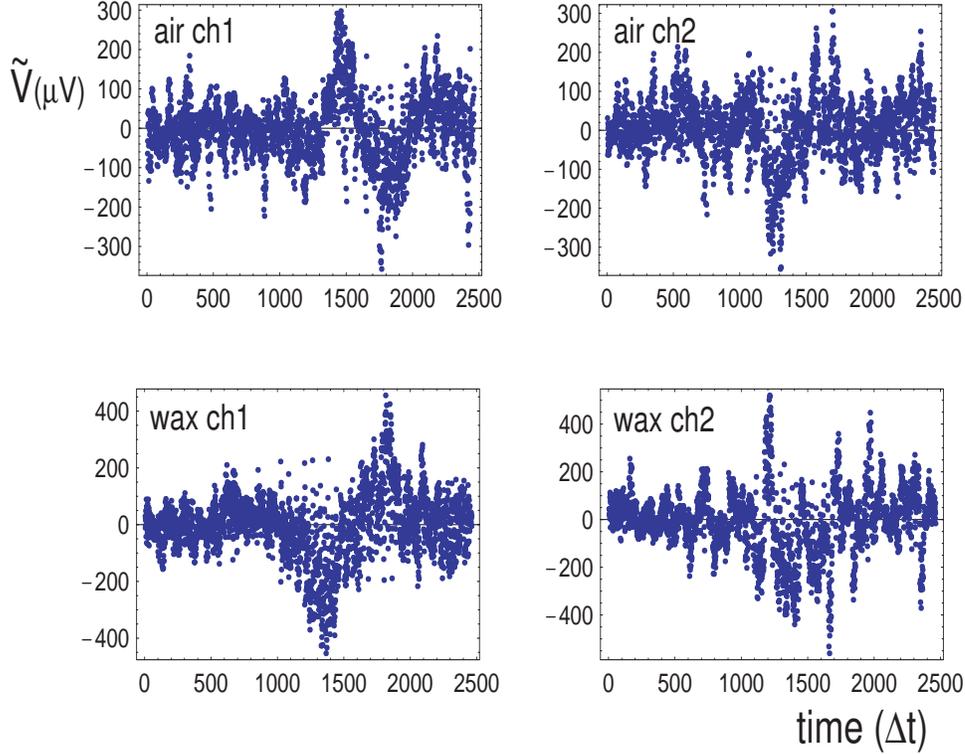}
\caption{ \small Output traces for filtered data sets $\tilde V$.  Free space medium (``air'', top) and wax (bottom), for
channel 1 (upstream, left panels) and channel 2 (downstream, right panels).  Causal onset can be seen in the signal region after the center of each dataset, as in Fig. \ref{fig:MCfliteredSignal}. Voltages in $\mu V$;  time in units of $\Delta t$=0.16ns.     }
\label{fig:bodbof}
\end{center}
\end{figure}

We examined the output $\tilde V(t)$ of averaged data for each dataset after filtering.  Fig. \ref{fig:bodbof} shows the filtered traces in two channels for the ``air'' and ``wax'' cases.  Recall that the air data is simply a beam traveling through empty space, and the wax data includes a lead pre-radiator that generates a hadronic shower.  Visual inspection shows a difference between the initial region and the region associated with causal arrival of the beam.  (The earliest causal onset point, as determined from our phototube and click-file calibration, is point 970 for channel 1.)  Since the signal to noise was very poor to begin with, we contented ourselves with simple statistical methods to quantify detection. 

\subsubsection{Data Distributions} 

\label{sec:datadistrib}

We divided the filtered data into two regimes, ``before'' the causal onset point (the first 970 points), and ``after'' (1490 points).  The total number of 2460 points comes from the filter retaining 77 bins of 32 points/bin.  The standard deviation $\sigma$ of the data sets before and after onset are recorded in Table 1.  Several sets show hundreds of points in the post-onset region with amplitudes larger than 3$\sigma_{before}$ (Table 1).  These statistics are evidence for events related to the arrival of the beam in both the air (virtual radiation) and wax (real radiation) cases.

\begin{table} [!]
  \centering 
  \begin{tabular}{ccccc}
\hline
% after \\ : \hline or \cline{col1-col2} \cline{col3-col4} ...
$channel $   & $ \sigma_{before} \, (N_{before})$  &   $ \sigma_{after} \, (N_{after})$ & $ N(\tilde V_{before}>3  \sigma_{before} )$ &  $ N(\tilde V_{after}>3  \sigma_{before}) $ \\
\hline
Air Ch 1 & 59\, (970)&   105 \, (1460) & 10  &  166 \\
Air Ch 2  & 74\, (970) &  103 \, (1460)  & 16 &    91 \\
Wax Ch 1  & 67\, (970)  & 150  \, (1460)   & 8  & 284    \\
Wax Ch 2  & 81\, (970) &  170 \, (1460)   & 6  &   266 \\
\hline \\
\end{tabular}
\caption{\small Basic statistics of the filtered data signal $\tilde V$ before causal onset and after.  Standard deviations ($\sigma$, in units of $\mu$V), number of points ($N$), and number of points exceeding 3$\sigma$ in the four channels measured.  \medskip
}

\end{table}

\begin{figure}
\begin{center}
\includegraphics[width=5in]{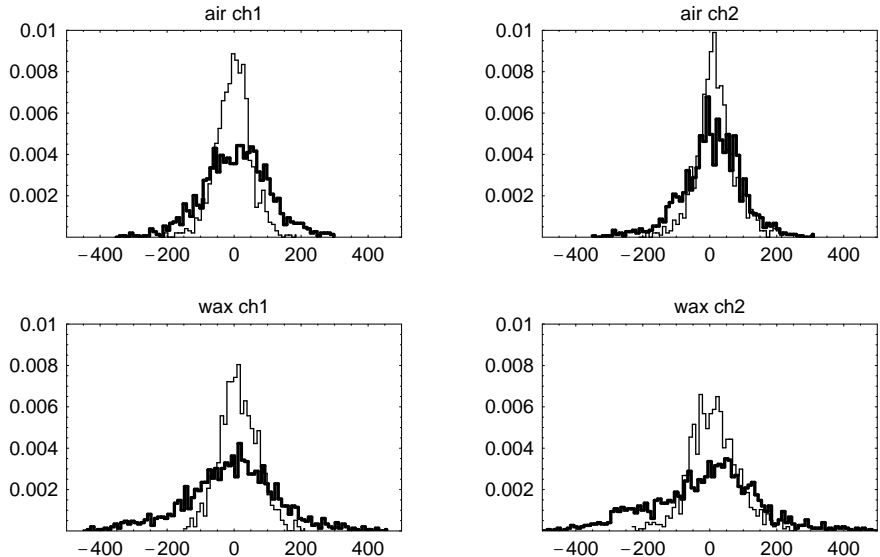}
\caption{\small Histograms of filtered data $\tilde V$ in $\mu$V from the time segments {\it before} causal onset (light lines) and {\it after} causal onset (heavy lines), see text.  The four panels are for the cases of free space medium (``air'', top) and wax (bottom), for channel 1 (upstream, left panels) and channel 2 (downstream, right panels). }
\label{fig:4histos}
\end{center}
\end{figure}

More information is given in Fig. \ref{fig:4histos}.  The figure compares histograms of the filtered data in the before and after regions. The distributions are reasonably consistent with Gaussian forms, but there is insufficient statistics to address the crucial issue of behavior in the tails. The histograms support the information given by the standard deviations, namely that the distributions after onset are wider than before.  We estimated $P$-values for the data before onset to fluctuate to the degree seen after onset.  If one assumes Gaussian distributions the $P$-values for all four cases were less than $10^{-9}$.  $P$-values depend on the number of degrees of freedom, for which we used 20/32 of the number of points to account for the reduction in freedoms from filtering.

\subsubsection{Time Structure} 
\label{sec:timestucture}

The statistics just cited are ``bulk'' integrated measures that contain no information on the time-structure of the data.  We attempted to quantify agreement of the shape of the filtered data with the filtered simulation, as follows. 

Recall that $\tilde V_{t}^{signal}$ is the filtered simulation, and $\tilde V_{t}$ is the filtered data. Remove the means and normalize each vector. Then $C(0)=\sum_{t} \, \tilde V_{t}\tilde V_{t}^{signal}$ represents the dot product of the two vectors, which is a measure of agreement of the shapes. Perfect agreement corresponds to $C(0)=1$, which is statistically very unlikely for vectors with many components.  More generally, one can shift one pattern relative to another by defining running correlations $C(T)=\sum_{t} \, \tilde V_{t+T}\tilde V_{t}^{signal}$, which happens to coincide with the 2-point correlation of classical statistics.  

The width of maximum $C(T)$ depends on the detailed way in which it is calculated.  For one thing, a very extended pattern degrades timing information, while tending to increase statistical significance.   Generally our studies yielded a timing resolution of order 200 sample time units (32 ns), which was consistent with our simulation.  The timing resolution after filtering is too poor to resolve the time separation between the two antennas to provide evidence of causality.  However the observation in the causal onset region can only come due to fields traveling with the beam.   Observation of 2 traces with consistent timing, shape, and size predicted by simulations provides evidence that the experiment detected virtual radiation.

We constructed our simulation in the simplest way possible because we have minimal information on the fluctuations of the beam position event-to-event, and also the bucket structure occasionally changed significantly. Rather than throw away data it was to our advantage to make averages. When we made running correlations $C(T)$ of our filtered data with the simulations, we found evidence supporting good agreement of the simulation time structure and data time structure in the form of large correlations $C(T) \gtrsim 0.4-0.5$ for simulation lengths of 800 points.  This was seen in each of the 4 data sets, and similar results were seen for many different simulation lengths.  The naive statistical probability one might estimate for the results in a random sample were vanishingly small in many cases.  At the same time, we found non-negligible correlations of some data sets in regions of $T$ that were supposed to correspond to noise. This is explained in part by the fact that the filter introduces correlations by retaining a class of data patterns that is not ideally random.  
The running correlations studies also depend on the length of the simulation region, the bin dimension and rank $D, \, R$, and the length of offset $T$, making for a very complicated statistics problem.  For this reason, we did not pursue a full statistical analysis of $C(T)$ further. 

This analysis gives evidence for detection based on appearance of signals in the causal region.  Noise fluctuations have been ruled out. There remains only a possible RF background, which if postulated must be closely timed event by event, and also pass the filter.  The Fourier spectrum of the phototube pulse was measured and is predominantly below 100 MHz, explaining our 200 MHz high-pass filter choice. Lack of complete immunity to phototube noise was clear in ``whopper'' events when the beam was steered inadvertently into the phototube itself.  These events were thrown out early.  However, we also quantified phototube noise with runs 
with finger-counter triggers to eliminate this possibility of our trigger signal polluting our antenna data.  

To quantify the signal further we turn to evidence of coherence of the signal. 

\section{Coherent Scaling} 

The hallmark of coherent Cherenkov radiation is linear scaling of the signal with the total number of particles. To test linearity one might vary the total number of charged particles in the beam.  However it is unrealistic to ask the test beam facility to adjust particle number while maintaining a consistent time-structure.  We varied the particle number by adjusting the number of runs included in the data analysis. 

\begin{figure}[htb]
\begin{center}
\includegraphics[width=5in]{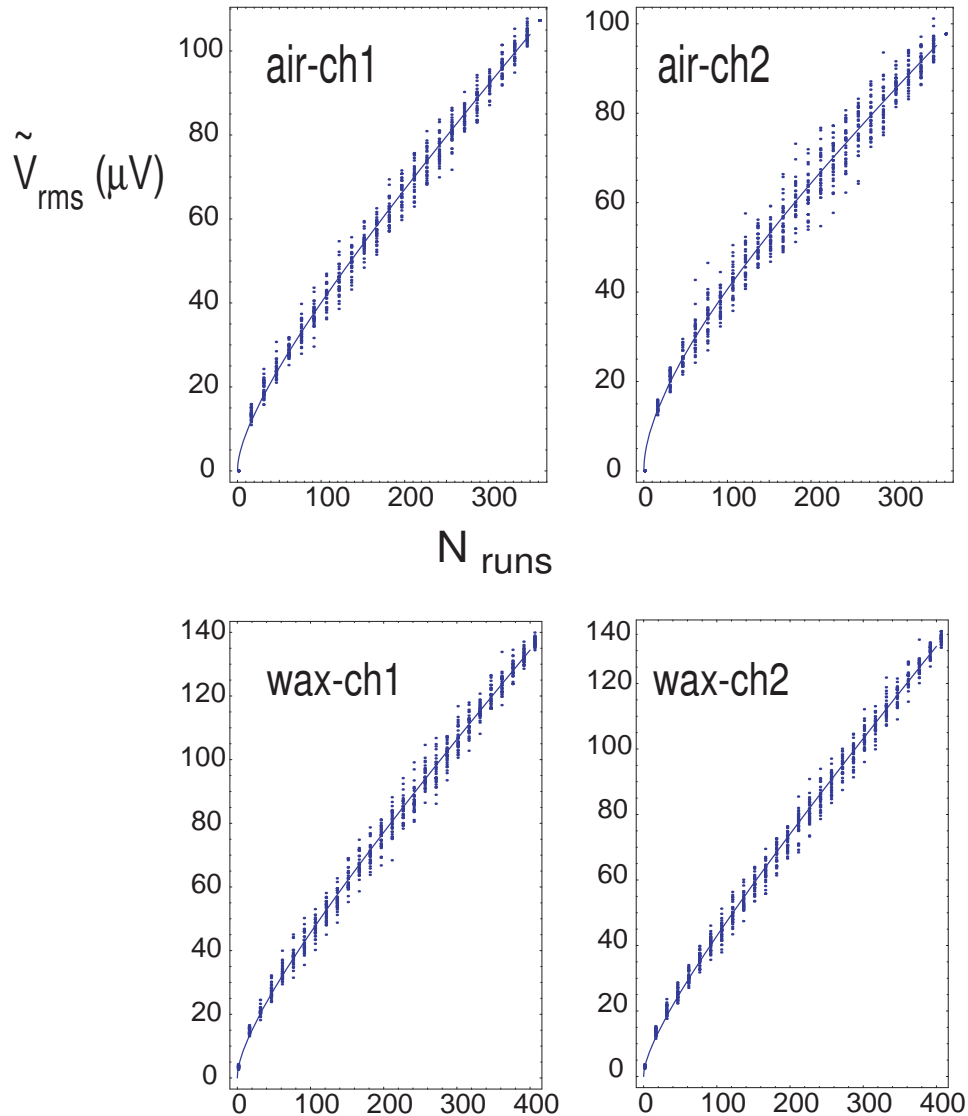}
\caption{ \small Dependence of $rms$ filtered voltage $\tilde V_{rms}$ on the number of particles, as represented by the number of randomly-permuted runs $N_{runs}$ analyzed.  Data from the {\it signal} region, as defined in the text. Curves are $\tilde V_{rms}(N_{runs}) =\a N^{1/2}$ and $\b N$, with parameters given in Table \ref{tab:sigfits}.}
\label{fig:4Powers.eps}
\end{center}
\end{figure}

We sum the filtered voltages for a given number of runs $N_{runs}$. We then calculate the $rms$ of the result, producing \ba \tilde V_{rms}(N_{runs}) =rms( \,  \sum_k^{N_{runs}}  \, \tilde V_{k}, \,), \nn \ea where $k$ is the run number and symbol $rms$ takes the standard deviation.  The results of adding a subset of runs depends on the order in which runs are selected.  To make an unbiased sample we repeated everything 30 times with the order of runs randomly permuted.  Fig. \ref{fig:4Powers.eps} shows the results using the {\it signal region} (the last 1500 points) for the four cases of channel 1, 2 in Air and Wax.  The figures show evidence of linear scaling, {\it i. e.} coherence, both in the Wax and Air cases.

\begin{table}[t!]
  \centering 
  
  \begin{tabular}{ccc}
\hline
% after \\ : \hline or \cline{col1-col2} \cline{col3-col4} ...
 $channel$  & $\a $ & $\b$    \\
   
   \hline
% after \\ : \hline or \cline{col1-col2} \cline{col3-col4} ...
Air Ch 1   & 2.2  &  0.18  \\
Air Ch 2   &   2.9  &  0.12  \\
Wax Ch 1   &  2.4 &  0.22 \\
Wax Ch 2   &   2.0  &   0.23  \\
\hline
\end{tabular}
  \caption{ \small Parameters of the fit $V_{rms}(N_{runs}) =\a  N^{1/2}_{runs}  + \b  N_{runs} $ developed in the signal region, as defined in the text. Units of $\a$ and $\b$ are $\mu$V.  Correlated errors are discussed in the text. }
  
  \label{tab:sigfits}
\end{table}

We fit dependence of the $\tilde V_{rms}(N_{runs})$ on $N_{runs}$ to an ansatz \ba \tilde V_{rms}(N_{runs}) =\a  N^{1/2}_{runs}  + \b  N_{runs} . \label{fit} \ea We call parameter $\a$ the {\it random walk} component, representing scaling with $ N^{1/2}_{runs}$ typical of noise.  One may interpret $\a$, which has units of $\mu V$, as a filtered noise-standard deviation, subject to fitting uncertainties.  Parameter $\b$ is the {\it linear coherence} component. This parameter can be interpreted as the filtered electric field per 500-proton run, measured in $\mu V$ via pre-factors that include antenna response and amplifier gains.  For reference, our filtered simulation predicted a value of $\b \sim 0.1 \mu $V/run, with large systematic uncertainties already cited. 

Basic expectations for the $\a$ and $\b$ parameters are developed from Table 1. Air Ch 1, for example, has $rms$ voltages of $\sigma_{before}= 59 \,\mu$V in the region before causal onset, and $\sigma_{after}=105 \mu$V after causal onset.  These figures represent the cumulative outcome of averaging 361 files.  Assuming $\sigma_{before}$ is nothing but random walk noise predicts $\a_{before}\sim 59 \,\mu $V$ / \sqrt{361} \sim  3.1 \mu$V.  Naively subtracting this from the signal region after onset, and assuming the balance is due to linear coherence, predicts $\beta_{after} \sim 0.15 \mu $ V.  

Quantitative fits to the $N_{run}$ dependence are shown in Fig. \ref{fig:4Powers.eps}. The first and last cases of $N_{runs}$ have trivial fluctuations and were dropped, a small effect.   Table \ref{tab:sigfits} shows the linear coherence and random walk components are very similar for the four data sets.   The fits are quite consistent with the simple argument given above.  Fitting two similar powers such as $\a N^{1/2}+\b N$ is known to be ``ill-conditioned.''  Each parameter can simulate the effects of the other over a finite $N$ range.  For this reason the fit parameters need to be assessed with correlated errors.  Discussion of the error ellipses is given in Section \ref{sec:contours}.

Thus the evidence for signal observation for the {\it averaged} data sets cited in Section \ref{sec:datadistrib} is consistent with {\it varying} the number of protons in the beam,  extracting the term linear in the number of protons, and then averaging.

\subsection{Errors on Coherent and Incoherent Contributions}
\label{sec:contours}

We now describe a more detailed study that develops an error estimate of parameters $\a$, 
$\beta$, which will be used in calculating a $\chi^{2}$ goodness-of-fit measure, described below. 

Recall that Fig. \ref{fig:Bod1Coherence.eps} cited earlier shows fits for $\tilde V_{rms} = \a  N^{1/2}_{runs}  + \b  N_{runs}$.  The top data set and curves (blue online) comes from the signal region, while the bottom data and curves (red online) come from the noise region. 
The random walk component is the only significant term in the noise region.  This is consistent with the observed difference of noise and signal region established in Section \ref{sec:datadistrib}.  It is also consistent with the random walk contribution of the signal region, Table \ref{tab:sigfits}.  A very small, negative (unphysical) linear term is generated by the best-fit procedure.

\begin{figure}[htb]
\begin{center}
\includegraphics[width=4in]{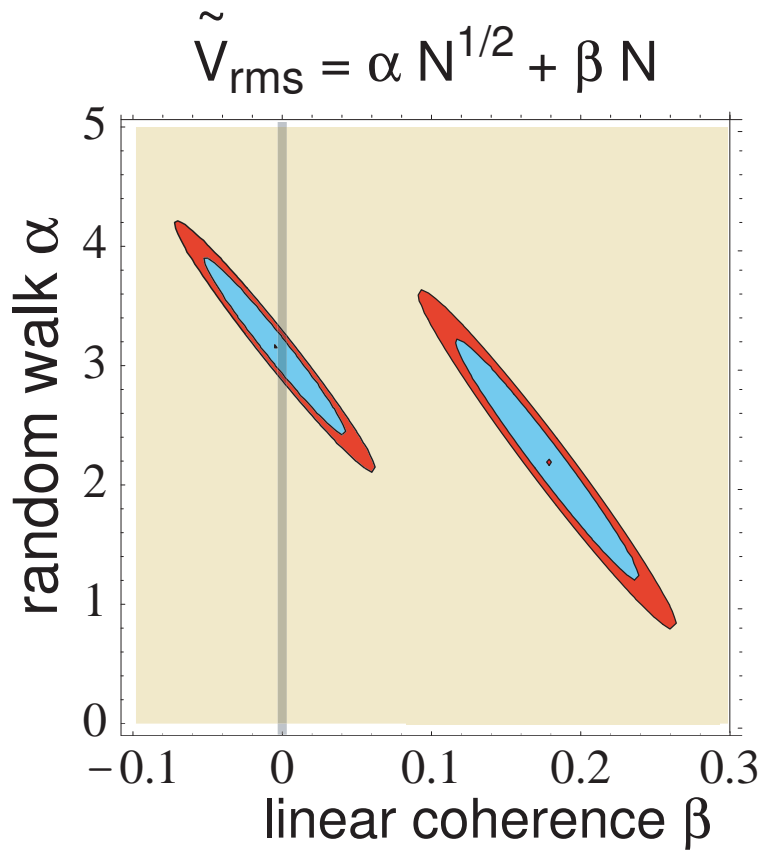}
\caption{ \small Contour plot of $\chi^{2}$ of the signal region and noise region versus parameters $\a, \, \b$.  Contours represent unit intervals of $\chi^{2}$.  The minimum $\chi^{2}/dof$ (central dots) is close to one.  }
\label{fig:cohjoinfit.eps}
\end{center}
\end{figure}

Each data set has comparable correlated errors. We present the details of analysis of 
Air Ch 1.  Fig. \ref{fig:cohjoinfit.eps} illustrates contours of goodness of fit versus parameters $\a, \, \b$.  Goodness of fit for this figure is defined by the usual $\chi^{2}$ formula, using the average of the fluctuations seen in Fig. \ref{fig:Bod1Coherence.eps} as the statistical error.   Dependence on $N_{runs}$ of $\tilde V_{rms}$ is separated into using the {\it signal region} (the last 1500 points) and the {\it noise region} (the first 900 points).  

In principle Wax Ch 1 and Air Ch 1 should differ primarily due to the lead pre-radiator and radio cavity mode re-configuation from adding the dielectric.  If hadronic fluctuations were a substantial effect we might see larger signals in the Wax case for both Ch 1 and Ch 2.  The nominal Cherenkov channel is Wax Ch 2, which in theory should have smaller signals due to the factor of $1/\epsilon$ in Eq. \ref{solution}.  We observe that Air Ch 2 has a somewhat lower $\beta$ parameter than the others. It is interesting that the linear coefficients $\beta$ in Table \ref{tab:sigfits}) for Wax (Cherenkov radiation) are somewhat larger than those in Air (virtual radiation).  We were gratified that the different cases are so comparable.  First, radio-frequency normalizations are very difficult to control to the level of a few $dB$. Second, the response of the tank at the location of the antennas depends on the presence of the dielectric.  The antenna for Ch 2 was also re-positioned and re-connected between runs.   The uncertain accumulation of physical effects degrades our ability to accurately verify normalizations.  For these reasons, and the ill-conditioning of power-law fits, we find the consistency of  parameters seen in Table 2 quite acceptable.

The random walk/linear coherence parameter regions of the signal and noise region are well separated.  As expected each fit has a substantial degeneracy in $\a, \, \b$ parameters.  Good fits to the signal region are obtained over a line $\a \sim 5.05 - 15.7 \b$, for $0.08<\b<0.22$.  Despite the degeneracy of the fit, there is no overlap of the error ellipse of the noise region with the signal region in Fig. \ref{fig:cohjoinfit.eps}.  This demonstrates that the signal region can only be fit with a dominant linear coherence behavior.

\subsection{Signal Shape} 

\label{sec:shape}

\begin{figure}
\begin{center}
\includegraphics[width=4in]{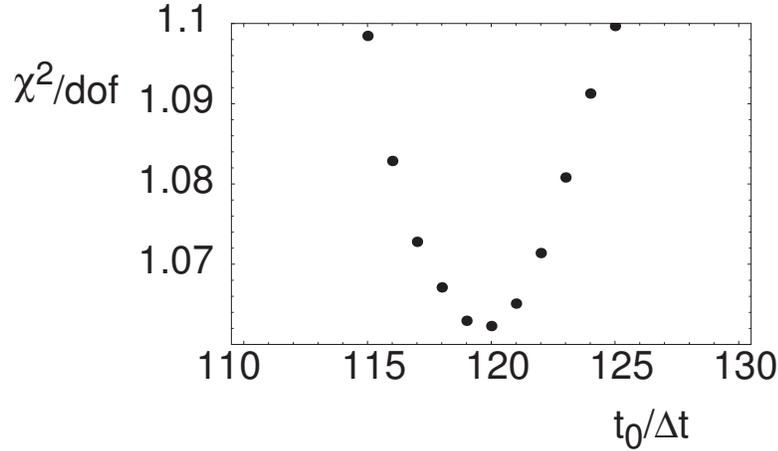}
\caption{ \small Goodness of fit $\chi^{2}/dof$ as a function of the timing offset
parameter in units of the sample time. }
\label{fig:ChiSqBod1Shif.eps}
\end{center}
\end{figure}

Finally we quantified the shape of the filtered data set (average of all runs) compared to the filtered simulation.  We gave the simulation 2 parameters, consisting of a normalization parameter $n$ and timing offset $t_{0}$: \ba \tilde V_{fit}(t)= n  \tilde V_{sim}(t-t_{0}). \nn \ea The parameter $n$ accounts for uncertainties in the normalization of the simulation and the number of particles in the beam spill. The timing offset $t_{0}$ accounts for uncertainties in timing delays, the bucket form factor, and in extra timing offset caused by the binning procedure, which introduced the discrete time bins used to make the filters.  

\begin{figure}
\begin{center}
\includegraphics[width=4in]{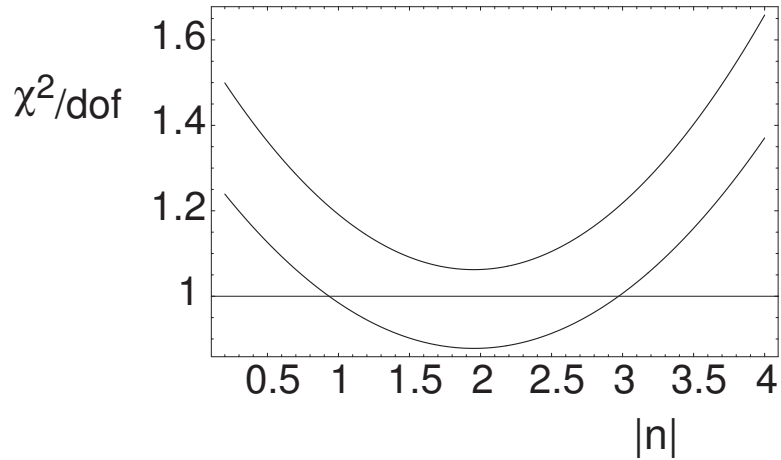}
\caption{ \small Goodness of fit $\chi^{2}/dof$ as a function of the magnitude parameter $|n|$.  Top curve uses $\a= 2.24$, bottom curve $\a=2.44$.  Both curves are acceptable due to degeneracy in estimating $\a$ (Fig. \ref{fig:cohjoinfit.eps}) . The statistical ideal of $\chi^{2}/dof=1$ is the horizontal line.    }
\label{fig:ChiSquareBod1.eps}
\end{center}
\end{figure}

The procedure is illustrated for Air Ch 1.  The results of the other channels are comparable within uncertainties. Specifically, Air Ch 1 is the case best predicted, while the analysis of the other channels differ by less than a factor of 2 in $\chi^{2}$ that can be attributed to the errors.  The best fit to the timing offset parameter (Fig. \ref{fig:ChiSqBod1Shif.eps}) is about 120 sample time units. Recall that the filter bin dimension is $D=32$ putting the onset around data point 1340. The timing offset is about 4-bin sizes of delay.  The offset delay is consistent with model studies in which a few bins in a signal region are needed to develop significant filtered signal/noise. The best fit of the magnitude parameters is -1.95, indicating a polarity was reversed.  A plot of $\chi^{2}/dof$ versus $|n|$ is shown in Fig. \ref{fig:ChiSquareBod1.eps}. The number of degrees of freedom $dof$ are the number of points in the file (2460) minus the number of parameters (2).  For these plots the estimated statistical fluctuation (``$\sigma$'') in the denominator of $\chi^{2}$ uses the central value of the random walk parameter $\a =2.24$ (Fig.\ref{fig:cohjoinfit.eps}). The minimum $\chi^{2}/dof$ for the central value of $\a$ is about 1.06, which is perfectly acceptable. 

Due to $\a, \,\b$ parameter degeneracy, the random-walk parameter is not well fixed to the central value, and can vary without substantially changing the goodness of fit.  From Fig. \ref{fig:cohjoinfit.eps} the $\a$ parameter can be varied by about 50\%, which suffices to account for variations of signal and noise observed between channels.  A decrease of $\a$ by a mere 10\%  is statistically insignificant and causes the best fit $\chi^{2}$ value to drop below 0.9 (Fig. \ref{fig:ChiSquareBod1.eps}, bottom curve).  Assuming $\chi^{2}/dof$ should be near 1 on general statistical grounds, an error on parameter $\b$ can be assigned; the result is $\b \sim 0.18  \ra \ 0.21 $, a relative change of order 15\%, which stays well within the error ellipse of Fig. \ref{fig:cohjoinfit.eps}.

\section{Summary}

We conducted an experiment seeking to measure the impulsive fields from 120 GeV protons passing radio antennas in a sealed environment.   A full system simulation was constructed describing charges moving at subliminal velocities in air (virtual radiation) and in wax (real Cherenkov radiation). One set of data simply measures propagation in air inside the tank.  In the second dataset we observed signals consistent with real Cherenkov radiation in wax after beams passed through a lead preradiator.  It is difficult to distinguish experimentally between the two experimental setups under the very near-field conditions of the apparatus.  Both experiments registered signals in the causal onset region consistent with the virtual fields of charges moving at subluminal velocities.  The signals demonstrate linear coherence and are consistent with simulation. We believe this is evidence for the first direct observation of radio frequency pulses from hadronic showers.  The sizes of the two sets of signals were comparable and matched theoretical expectations.

A method extracting signal from data with very small signal-to-noise ratio has demonstrated novel and highly effective filtering techniques.  The techniques are promising and may be applied broadly to improve experimental analysis of conditions with dramatically low signal to noise

Acknowledgments:  We greatly appreciate support from Fermilab personnel, especially Eric Ramberg and Chuck Brown, whose dedicated help was indispensable. For wax casting we also thank Bob Wertz, Waxman Candles Inc., Lawrence KS and Chicago IL.   Work supported under Kansas-DOE-EPSCOR-KASP project, DOE grant number DE-FG02-04ER14308, KU-GRF and KU-ERC.

\end{document}